\documentclass[useAMS,usenatbib]{mn2e}
\usepackage{graphicx}

\title[Multi-site campaign on KIC\,6382916]
{Multi-site photometric campaign on the high amplitude $\delta$\,Scuti star KIC\,6382916}
\author[Ulusoy et al.] 	
{C. Ulusoy$^{1,2}$, B. Ula\c{s}$^{3}$, T. G\"{u}lmez$^{1}$, L. A. Balona$^{4}$, I. Stateva$^{5}$, I. Kh. Iliev$^{5}$,
\newauthor
D. Dimitrov$^{5}$, H. A. Kobulnicky$^{6}$, T. E. Pickering$^{4,7}$, L. Fox Machado$^{8}$,
\newauthor 
 M. \'{A}lvarez$^{8}$, R. Michel$^{8}$, K. Antoniuk$^{9}$, D. N. Shakhovskoy$^{9}$, N. Pit$^{9}$,
\newauthor  
M. Damasso$^{10,11}$, D. Cenadelli$^{11}$, A. Carbognani$^{11}$ \\
\\
$^1$Department of Physics, University of Johannesburg,  P.O. Box 524, APK Campus, 2006, Johannesburg, South Africa\\
$^2$Department of Physics, Izmir Institute of Technology, 35430, Izmir, Turkey\\
$^3$Department of Astronomy and Space Sciences, University of Ege, 35100, Bornova, Izmir, Turkey\\
$^4$South African Astronomical Observatory, P.O. Box 9, Observatory 7935, 
Cape Town, South Africa \\
$^5$Institute of Astronomy with NAO, Bulgarian Academy of Sciences,
blvd.Tsarigradsko chaussee 72, Sofia 1784, Bulgaria\\
$^6$Department of Physics $\&$ Astronomy, University of Wyoming, Laramie, WY,
82071, USA\\
$^7$Southern African Large Telescope Foundation, P.O. Box 9, 7935 Observatory, Cape Town, South Africa\\
$^8$Observatorio Astron\'{o}mico Nacional, Instituto de Astronom\'{i}a, Universidad Nacional Aut\'{o}noma de M\'{e}xico,\\
Apartado Postal 877, 22830, Ensenada, B.C., M\'{e}xico\\
$^9$Crimean Astrophysical Observatory, Scientific Research Institute, 98409, Nauchny, Crimea, Ukraine\\
$^{10}$Astronomical Observatory of the Autonomous Region of the Aosta Valley (OAVdA) Loc. Lignan 39, 11020 Nus (Aosta), Italy\\
$^{11}$Department of Physics and Astronomy, University of Padova, vicolo dell'Osservatorio 3, I-35122 Padova, Italy\\
}

\begin{document}

\date{ .... Received ...}

\pagerange{\pageref{firstpage}--
\pageref{lastpage}--\pubyear{2013}} 

\maketitle
\label{firstpage}

\begin{abstract}
We present results of a multi-site photometric campaign on the high-amplitude 
$\delta$\,Scuti star KIC\,6382916 in the {\it Kepler} field.  The star was 
observed over a 85-d interval at five different sites in North America and 
Europe during 2011. {\it Kepler} photometry and ground-based multicolour light 
curves of KIC\,6382916 are used to investigate the pulsational content and
to identify the principal modes. High-dispersion spectroscopy was also obtained 
in order to derive the stellar parameters and projected rotational velocity.  
From an analysis of the {\it Kepler} time series, three independent frequencies 
and a few hundred combination frequencies are found.  The light curve is 
dominated by two modes with frequencies $f_{1}$= 4.9107 and $f_{2}$=
6.4314\,d$^{-1}$. The third mode with $f_{3}$= 8.0350\,d$^{-1}$ has a much lower 
amplitude.  We attempt mode identification by examining the amplitude 
ratios and phase differences in different wavebands from multicolour photometry
and comparing them to calculations for different spherical harmonic degree,
$l$. We find that the theoretical models for $f_1$ and $f_2$ are in a best agreement
with the observations and lead to value of l = 1 modes, but
the mode identification of $f_3$ is uncertain due to its low amplitude. 
Non-adiabatic pulsation models show that frequencies below 6\,d$^{-1}$ are
stable, which means that the low frequency of $f_1$ cannot be reproduced.
This is further confirmation that current models predict a narrower
pulsation frequency range than actually observed.
\end{abstract}

\label{firstpage}
\section{Introduction}

The high--amplitude $\delta$ Sct (HADS) stars  are defined as a Population I
subgroup of $\delta$\,Sct type variables. They are located in the central part of 
the instability strip \citep{mc00} in the core or shell hydrogen burning stage of 
stellar evolution and appear to be intermediate between normal
$\delta$~Scuti stars and classical Cepheids.  However, the 
distinction between HADS and other $\delta$\,Sct stars is still rather 
arbitrary \citep{soz08}.  Their large amplitude, which typically exceeds
0.3\,mag, and the presence of many combination frequencies caused by nonlinear 
coupling between the principal modes are the defining characteristics \citep{bre98}.  
In general, the HADS appear to pulsate mostly in the fundamental and first 
overtone radial modes \citep{bal12a}. However, they need not to be purely 
radial pulsators, since recent studies have shown that some high-amplitude
modes are nonradial \citep{mc00, por03, por11}. 

KIC\,6382916 (ASAS 194803+4146.9, $V= 10.79$) was first monitored during the 
ASAS--3 North station observations and reported as a double-mode HADS star with
a period ratio $P_1/P_0= 0.763$ by \cite{p09}. The star has also been observed by 
the  {\it Kepler} satellite in short-cadence (SC, 1--min exposures) and long-cadence 
(LC, 30--min exposures) modes \citep{gil10, jen10}. The  {\it Kepler} observations 
are very important as they allow us to fix the frequencies with great precision.  
We can use these frequencies to fit the multicolour ground-based observations 
and to determine the amplitudes and phases for the purpose of mode identification.

In this paper, we present results of a multi-site photometric and spectroscopic 
campaign on the HADS star KIC\,6382916 in order to identify the modes of
pulsation.  Mode identification is the first step in using the frequencies
to determine the stellar parameters (asteroseismology).  The paper is structured 
as follows: we first present a detailed description of the ground- and space-based 
observations including method of data reduction and frequency analyses. Mode 
identification, which is the main purpose of this study, is presented in Section 6. 
Finally, we discuss these results.

\section{Spectroscopy}

Spectroscopic observations were obtained at two different sites.  The first
set of spectra were obtained with the 2-m RCC telescope of the Bulgarian
National Astronomical Observatory, Rozhen. We observed the star during two
nights (2011 July 8 and 9) and in three spectral regions 4800--4910\,\AA\ 
(H$\beta$), 4500--4610\,\AA\ (Si lines) and 6390--6500\,\AA\ (Fe lines). A 
Photometrics AT200 camera with a SITe SI003AB $1024 \times 1024$ CCD chip
($24\,\mu {\rm m}$ pixel size) was used in the third camera of the Coud\'e spectrograph 
to provide spectra with a typical resolution of R = 32\,000 and a
signal-to-noise (S/N) ratio of about 50. The exposure times were 1800\,s. 
The intrinsic spectral line profile, which halfwidth gave about 9\,km\,s$^{-1}$near
6500\,\AA, was determined from the arc spectrum. Standard IRAF procedures 
were used for bias subtraction, flat-fielding and wavelength calibration. 

We also obtained spectra of KIC\,6382916 using the WIRO longslit spectrograph
with an E2V $2048 \times 2048$ CCD as detector. An 1800~l~mm$^{-1}$ grating in first order 
yielded a spectral resolution of 1.5\,\AA\ near 5800\,\AA\ with a 1.2\arcsec$\times$100\arcsec\ 
slit.  The spectral coverage was 5250--6750~\AA. Individual exposure times were 600~s. 
Reductions followed standard longslit techniques. Each spectrum was shifted by a small 
amount in velocity so that the Na~I~D $\lambda\lambda$\,5890, 5996\, lines were 
registered with the mean Na\,I line wavelength across the ensemble of observations. 
This zero-point correction to each observation is needed to account for the effects 
of image wander in the dispersion direction when the stellar FWHM of the point 
spread function is appreciably less than the slit width.  Multiple exposures were 
then combined, yielding a final S/N ratio typically in excess of 60 near 5800\,\AA. 

\section{Atmospheric parameters}

Model atmospheres were calculated using the {\tt ATLAS\,12} code.  The VALD atomic 
line database (\citealt{kprsw99}), which also contains \citet{Kurucz93} data, was used 
to create a line list for the synthetic spectra.  The {\tt SYNSPEC} code (\citealt*{hlj94}, 
\citealt{Krticka98}) was used to generate synthetic spectra adopting a microturbulence of 
2\,km\,s$^{-1}$.  The computed spectra were convolved with the instrumental profile 
(a Gaussian of 0.2\,\AA ~FWHM for the Coud\`e spectra and 1.5\,\AA ~FWHM for the WIRO 
spectra) and rotationally broadened to fit the observed spectrum.

\begin{figure}
\includegraphics[scale=1.1]{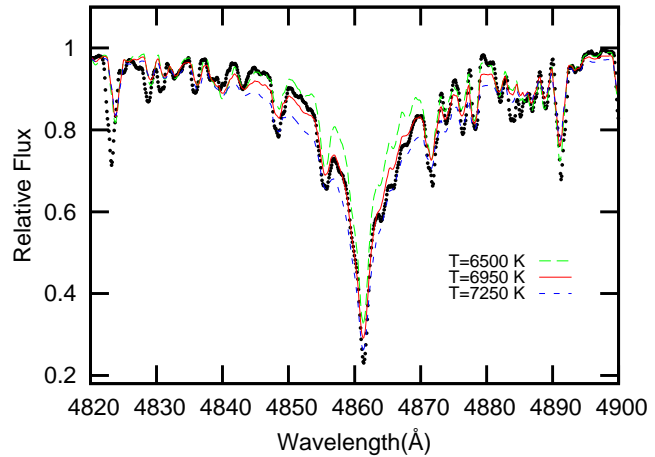}
\caption{H$\beta$ line (dots) fitted with a model with $T_{\rm eff}=6950~{\rm K}$, $\log g=3.7$ (cgs)
(solid line). Two other models - $T_{\rm eff}=6500~{\rm K}$ (long-dashed line) and $T_{\rm eff}=7250~{\rm K}$
(dashed line) are given for comparison.}
\label{figspec}
\end{figure}

The best fit for the H$\beta$ and H$\alpha$ lines was obtained 
for $T_{\rm eff}=6950\pm100~{\rm K}$,
$\log g=3.7\pm0.1$. We used the Mg\,II $\lambda$\,4481\,\AA\ line for the 
determination of projected rotational velocity.  The match between the synthetic 
and observed profile resulted in $v\sin\,i=50\pm10~{\rm km\,s^{-1}}$.   In
Fig.\,\ref{figspec} we show the best fit for H$\beta$ together with a fit
using two other effective temperatures for comparison.

\section{The {\it Kepler} Photometry}

{\it Kepler} data were used to derive the frequency content of KIC\,6382916. 
The {\it Kepler} Mission, designed to detect Earth-like planets using the 
transit method \citep{koh10}, was launched on 2009 March 6.  {\it Kepler}
has observed, and is continuing to observe, about 150\,000 stars in a fixed
field of view.  The superb photometric precision and the almost continuous 
data coverage is of great advantage in determining the pulsational frequencies 
which can then be used to fit the ground-based data.  {\it Kepler}
observations consist mostly of long-cadence (LC) exposures of 30-min
duration, but a few thousand stars, including KIC\,6382916, were observed
using 1-min (short cadence, SC) exposures.

The LC data are of limited value since the maximum frequency that can be 
extracted is about 24\,d$^{-1}$.  We therefore used only the SC data which 
consists of 38314 points between JD\,2455064.38 and JD\,2455091.48
(27.1\,d) taken at {\it Kepler} quarter 2.3 (Q2.3).  With SC data
frequencies as high as 700\,d$^{-1}$ can be detected if they are present.
The data were prepared for analysis by cotrending and detrending the 
Simple Aperture Photometry (SAP) fluxes. The cotrending process was applied to 
the Q2.3 data using Cotrending Basis Vector (CBV) files which help to remove 
instrumental systematics from the light curve \citep{fre12,chr12}. {\tt KEPCOTREND}\footnote{{\it  http://keplergo.arc.nasa.gov/ContributedSoftwareKepcotrend.shtml}} package that is provided by NASA Kepler Science Center is used during cotrending process. All data points were converted to magnitudes (m$_{i}$) using the formula $m_i =-2.5\log F_i$, where $F_i$ is the raw SAP flux.  A linear trend to $m_i$ as
a function of time was removed so that the final magnitudes have zero
mean.

\begin{figure}
\centering
\includegraphics{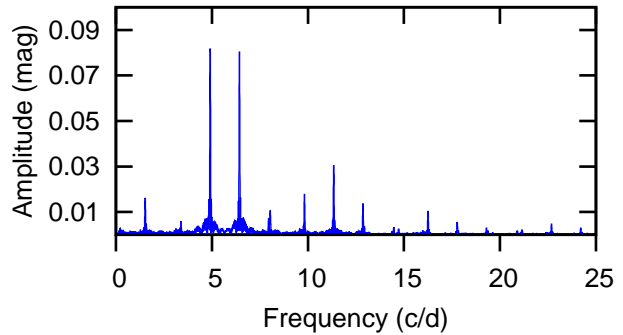}
\caption{Fourier spectrum of KIC\,6382916 for SC {\it Kepler} data }
\label{spectkep}
\end{figure}

We used {\tt PERIOD04} \citep{l05} to perform the frequency extraction.
Frequencies were extracted by successive prewhitening until the signal-to-noise 
threshold S/N $=$ 3.5 was reached.  All peaks with S/N greater than this
value were deemed significant \citep{bre11}.  We found that the light curve can 
be described by two independent frequencies, $f_{1} = 4.9107$ and $f_{2} = 
6.4314$\,d$^{-1}$, together with their harmonics and a few hundred combination 
terms.  A third independent frequency, $f_{3} = 8.0350$\,d$^{-1}$, is also
present but has a much lower amplitude.   The amplitude spectrum is shown in 
Fig. \ref{spectkep}.  Apart from $f_1$ and $f_2$, the peaks of highest
amplitude are the combination terms $f_1 + f_2$ and $f_2 - f_1$ and the 
harmonics $2f_1, 2f_2$.  The lowest frequency that appears significant is 
$f_2- f_1 = 1.520$\,d$^{-1}$.  For the calculation of phase differences and 
amplitude ratios at different wavebands, we only considered the first seven 
frequencies of highest amplitude.  These frequencies and their amplitudes and 
phases are listed in Table \ref{tabkep}.

\begin{table}
\caption{Modes of highest amplitude in KIC\,6382916 extracted from {\it Kepler} 
photometry.  The first column is the name of the mode. The frequencies $f$ (d$^{-1}$), 
amplitudes $A$ (mag) and phases $\phi$ (radians) and the signal-to-noise
ratio are listed.  The standard deviation in the last digits is given.
The epoch of phase zero is BJD\,2454833.00.}
\label{tabkep}
\begin{tabular}{lrrrr}
\hline
\hline
Name  &  \multicolumn{1}{c}{$f$ (d$^{-1}$)}&    \multicolumn{1}{c}{A (mag)} &   \multicolumn{1}{c}{$\phi$} &  \multicolumn{1}{c}{$S/N$} \\
\hline
$f_1$ 	            &	4.910722(8)   	&	0.08067(3)	&	0.26626(6)	&	1406\\
$f_2$ 	            &	6.431366(9)	&	0.07852(3)	&	0.39624(7)	&	1107\\
$f_3$              &	8.035055(54)	&	0.01233(3)	&	0.58971(42)	&	158\\
$f_4$ ($f_1+f_2$)   &  11.342088(22)	&	0.03077(3)	&	0.61926(17)	&	325\\
$f_5$ ($2f_1$)	    &	9.819598(39)	&	0.01734(3)	&	0.93893(30)	&	209\\
$f_6$ ($f_2-f_1$)   &	1.522490(34)	&	0.01624(3)	&	0.92359(27)	&	250\\
$f_7$ ($2f_2$)      &  12.863552(50)	&	0.01337(3)	&	0.23045(39)	&	144\\

\hline
\hline
\end{tabular}
\end{table}

\section{Ground-based Photometry}

Photometric observations of KIC\,6382916 were obtained at five different 
observatories located in North America and Europe.  A  total of 288 hours of 
observation was accumulated over 53 nights within the 85-d duration of the campaign.  
The campaign began in 2011 June and ended in 2011 September (JD 2455719.29--2455804.48). 
All data were obtained with either CCD or PMT detectors attached to five 
different telescopes in the ${\rm UBVRI}$ photometric bands. A detailed 
description of the observations is given in Table \ref{tabgb1}.

\begin{table}
\centering
\scriptsize
\caption{Information on the photometric multi-site campaign.  BNAO-Rozhen: 
Bulgarian National Astronomical Observatory-Rozhen; OAVdA: The Astronomical 
Observatory of the Autonomous Region of the Aosta Valley; SPM: Observatorio de 
San Pedro M\'{a}rt\'{i}r;  RBO: Red Butte Observatory; CrAO: Crimean Astrophysical 
Observatory.  The telescope apertures are in m. The total number of nights observed 
and the start end ending Julian dates of the observations relative to JD\,24455000 
are given.}
\label{tabgb1}
\begin{tabular}{lrrr}
\hline
\hline
Obs. &
Site &
Tel. &
CCD--PMT
\\
\hline
BNAO-Rozhen & Bulgaria & 0.60--0.70& FLI PL9000--16803 \\
OAVdA & Italy & 0.81 & FLI  PL3041-1-BB \\
SPM & Mexico  & 0.84 & FLI Fairchild F3041\\
RBO& USA      & 0.60 & Apogee Alta E47-UV \\
CrAO&Ukraine& 1.25 &FLI PL1001E-1\\
CrAO&Ukraine& 1.25 &5-channel UBVRI\\
         &           &         &photometer-polarimeter\\
\hline
Obs. &
Filters &
Dates &
Nights \\
\hline
BNAO-Rozhen & ${\rm BVRI}$     &  3 & 735.35--782.56 \\
OAVdA       & ${\rm BVRI}$      &  17 & 740.38--782.62\\
SPM         & ${\rm UBVR_cI_c}$ &  14 & 724.74--802.67\\
RBO         & ${\rm UBVRI}$      &  6 & 767.71--791.96\\
CrAO       &${\rm UBVR_cI_c}$&13&719.29--804.48\\
\\
\hline
\hline
\end{tabular}
\end{table}

\begin{figure}
\centering
\includegraphics{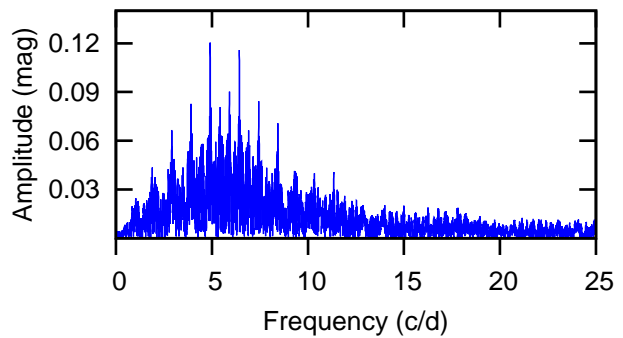}
\caption{Fourier spectrum of ground--based data before 
prewhitening}
\label{spectGB}
\end{figure}

The PMT data were only obtained using the 1.25-m telescope equipped 
with a five-channel photometer/polarimeter at CrAO. Since no $U$ filter was 
available at BNAO-Rozhen and OAVdA, CCD measurements were obtained only in 
$BVRI$.  Furthermore, the $U$ data from WIRO and CrAO suffers from
instabilities due to the poor quality of the data.  As a result, the frequencies 
in the $U$ band could not be properly resolved and therefore we decided to
omit the $U$ observations for the purpose of mode identification.  All the
data sets from different sites were checked carefully and prepared for 
Fourier analysis by removing outliers caused by bad weather conditions or 
some other reason.  The $R$ and $I$ filters are all of the Johnson/Bessel 
type except for SPM (Mexico) and CrAO (Ukraine) which are of the Cousins/Kron 
type.  These filter differences have a negligible effect on the resulting
amplitude ratios and the phase differences and no specific correction was
applied.

Data reduction was performed by following the standard procedures. For the 
CCD observations, the data were reduced with standard {\tt IRAF} routines 
including dark frame, bias subtraction and flat-field corrections for each CCD 
frame.  Data taken with the different instruments were reduced with respect 
to only one comparison star, GSC\,3144\,0053, which could be observed in the
CCD field at all sites. Instrumental magnitudes were obtained with the 
{\tt DAOPHOT II} package \citep{ste87} for aperture photometry. The PMT data 
from CrAO were reduced by following the traditional steps of obtaining differential 
magnitudes, sky subtraction and differential atmospheric extinction corrections 
using the same comparison star. Finally, the times were converted to 
Heliocentric Julian Date (HJD).

\begin{figure*}
\begin{center}
\centering
\includegraphics[scale=0.8]{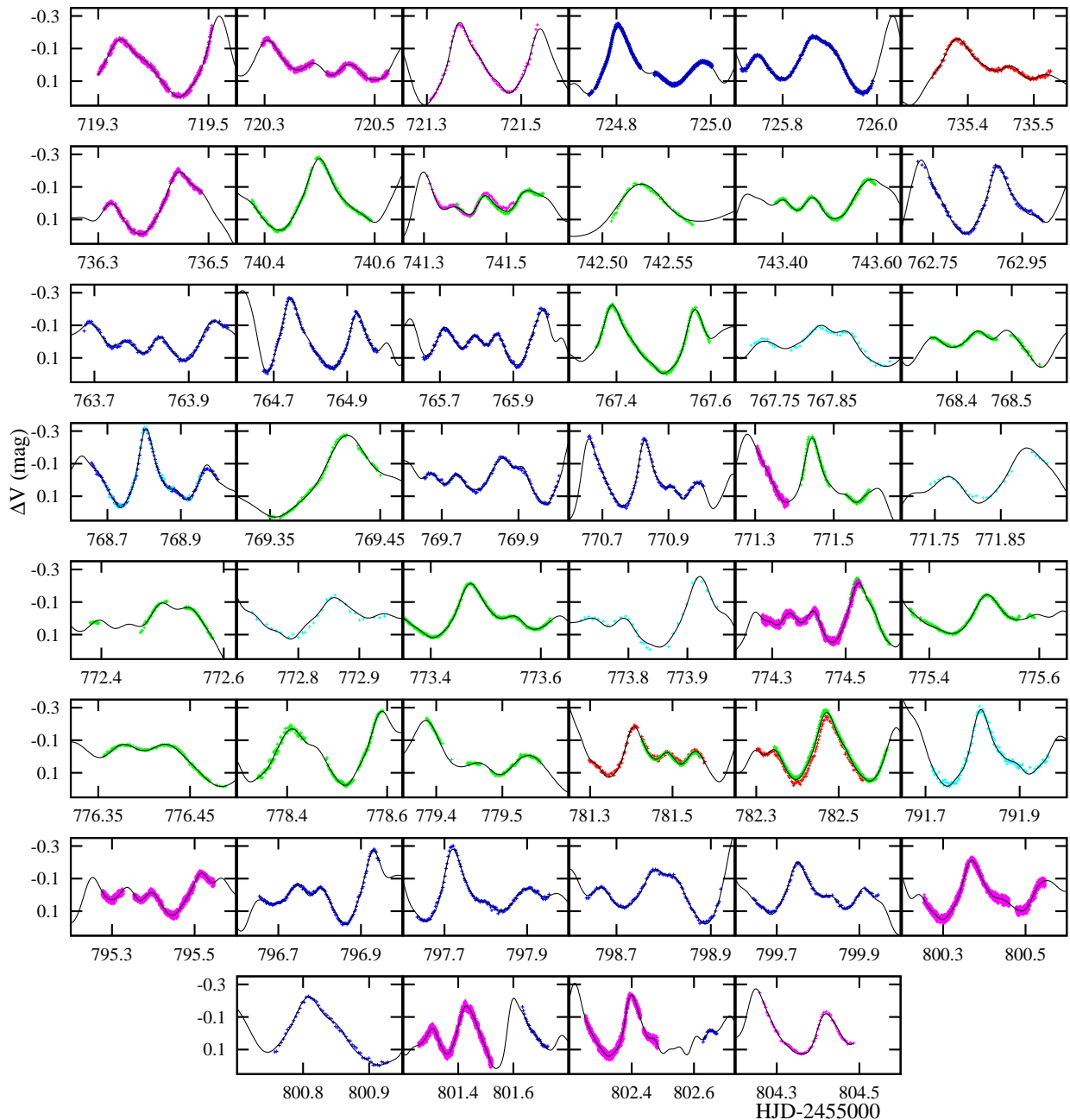}
\caption{V filter light curves corrected for zero points showing fitted Fourier curve.}
\label{figlcurve}
\end{center}
\end{figure*}

Since the telescopes and instruments used at the observing stations were not 
exactly the same, measurements from different sites have different zero points.  
In order to determine the zero point differences in each filter for each site, 
we assume that the light curve is well represented by a truncated Fourier
series using frequencies derived from the {\it Kepler} photometry \citep{ulu13}. 
The Fourier spectrum of the data before prewhitening can be seen in Fig. \ref{spectGB}. 
Using the first seven frequencies of highest amplitude extracted from the SC 
{\it Kepler} data, the individual zero points for the different sites together 
with the number of observations are shown in Table \ref{tabzeros}. The amplitudes 
and phases were determined by least-square fitting with the three independent 
frequencies: $f_1=4.9107$, $f_2=6.4314$ and $f_3=8.0350$\,d$^{-1}$, 
the harmonics of $f_1$ and $f_2$ ($2f_1$, $2f_2$) and the combination frequencies 
listed in Table. \ref{tabcoefs}. The uncertainities were calculated as the standard deviations of each fitted parameter. The observed light variation of the star in the 
$V$ filter together with the fitted Fourier series are shown in Fig. \ref{figlcurve}.

\begin{table}
\centering
\scriptsize
\caption{Zero points in the ${\rm UBVRI}$ filters for various sites and the number
of observations, $N$.}
\label{tabzeros}
\begin{tabular}{lrrrr}
\hline
\hline
\multicolumn{1}{c}{Site} &
\multicolumn{1}{c}{$U$} &
\multicolumn{1}{c}{$N$} \\
\hline
Mexico& $-2.863 \pm 0.001$ & $  1395$ \\
\hline
\multicolumn{1}{c}{Site} &
\multicolumn{1}{c}{$B$} &
\multicolumn{1}{c}{$N$} &
\multicolumn{1}{c}{$V$} &
\multicolumn{1}{c}{$N$} \\
\hline
Bulgaria & $-1.446 \pm 0.002$ & $ 373$  & $-0.276 \pm 0.001$ & $ 379$\\
Italy       & $-1.118 \pm 0.001$ & $  3474$& $-0.237 \pm 0.001$ & $ 3209$\\
Mexico   & $-1.289 \pm 0.001$ & $ 1435$ & $-0.288 \pm 0.001$ & $ 2814$\\
Ukraine  & $-1.255 \pm 0.001$ & $ 11795$& $-0.213 \pm 0.001$ & $ 9985$\\
USA         & $-1.211 \pm 0.002$& $   278$  & $-0.267 \pm 0.002$ & $   275$\\
\hline
\multicolumn{1}{c}{Site} &
\multicolumn{1}{c}{$R$} &
\multicolumn{1}{c}{$N$} &
\multicolumn{1}{c}{$I$} &
\multicolumn{1}{c}{$N$} \\
\hline
Bulgaria & $0.312 \pm 0.001$ & $ 372$  & $ 0.919 \pm 0.001$ & $  372$ \\
Italy       & $0.389\pm 0.001$  & $ 2922$ & $ 0.966\pm 0.001$ & $  2618$ \\
Mexico   & $ 0.317 \pm 0.001$& $  1420$& $ 0.942 \pm 0.001$ & $  1390$ \\
Ukraine  & $0.395 \pm 0.001$ & $ 10587$& $1.039 \pm 0.001$ & $ 9423$ \\
USA        & $ 0.361 \pm 0.002$   & $   273$  & $ 0.987\pm 0.001$ & $255$ \\
\\
\hline
\hline
\end{tabular}
\end{table}

\begin{table}
\centering
\scriptsize
\caption{Amplitudes, A (mag), and phases (radians) for the three frequencies 
of highest amplitude: $f_1=4.9107$, $f_2=6.4314$ and $f_3=8.0350$ d$^{-1}$, 
their harmonics and linear combinations of $f_1$ and $f_2$. The epoch of 
phase zero is JD 2455700.000.}
\label{tabcoefs}
\begin{tabular}{lrr}
\hline
\hline
\multicolumn{1}{c}{ID} &
\multicolumn{1}{c}{$A_B$} &
\multicolumn{1}{c}{$\phi_B$} \\
\hline
$f_{1}$	&	0.1281	$\pm$	0.0005	&	0.2668	$\pm$	0.0006	\\
$f_{2}$&	0.1263	$\pm$	0.0005	&	0.1294	$\pm$	0.0006	\\
$f_{3}$	&	0.0145	$\pm$	0.0005	&	0.4718	$\pm$	0.0056	\\
$f_{1}+f_{2}$	&	0.0469	$\pm$	0.0005	&	0.1292	$\pm$	0.0017	\\
$2f_{1}$&	0.0186	$\pm$	0.0005	&	0.3068	$\pm$	0.0043	\\
$f_{2}-f_{1}$	&	0.0139	$\pm$	0.0005	&	0.2081	$\pm$	0.0056	\\
$2f_{2}$	&	0.0226	$\pm$	0.0005	&	-0.3858	$\pm$	0.0036	\\

\hline
\multicolumn{1}{c}{ID} &
\multicolumn{1}{c}{$A_V$} &
\multicolumn{1}{c}{$\phi_V$} \\
\hline
$f_{1}$	&	0.0929	$\pm$	0.0003	&	0.2779	$\pm$	0.0006	\\
$f_{2}$	&	0.0925	$\pm$	0.0003	&	0.1228	$\pm$	0.0006	\\
$f_{3}$&	0.0141	$\pm$	0.0003	&	-0.4650	$\pm$	0.0037	\\
$f_{1}+f_{2}$	&	0.0335	$\pm$	0.0003	&	0.1219	$\pm$	0.0015	\\
$2f_{1}$&	0.0158	$\pm$	0.0003	&	0.3565	$\pm$	0.0033	\\
$f_{2}-f_{1}$	&	0.0150	$\pm$	0.0003	&	0.2216	$\pm$	0.0033	\\
$2f_{2}$	&	0.0144	$\pm$	0.0003	&	-0.3996	$\pm$	0.0036	\\

\hline
\multicolumn{1}{c}{ID} &
\multicolumn{1}{c}{$A_R$} &
\multicolumn{1}{c}{$\phi_R$} \\
\hline
$f_{1}$	&	0.0696	$\pm$	0.0003	&	0.2675	$\pm$	0.0006	\\
$f_{2}$	&	0.0683	$\pm$	0.0003	&	0.1157	$\pm$	0.0006	\\
$f_{3}$	&	0.0109	$\pm$	0.0003	&	-0.4611	$\pm$	0.0041	\\
$f_{1}+f_{2}$	&	0.0234	$\pm$	0.0003	&	0.1134	$\pm$	0.0019	\\
$2f_{1}$	&	0.0123	$\pm$	0.0003	&	0.3460	$\pm$	0.0036	\\
$f_{2}-f_{1}$&	0.0086	$\pm$	0.0003	&	0.1979	$\pm$	0.0050	\\
$2f_{2}$	&	0.0118	$\pm$	0.0003	&	-0.4180	$\pm$	0.0038	\\

\hline
\multicolumn{1}{c}{ID} &
\multicolumn{1}{c}{$A_I$} &
\multicolumn{1}{c}{$\phi_I$} \\
\hline
$f_{1}$	&	0.0487	$\pm$	0.0003	&	0.2578	$\pm$	0.0009	\\
$f_{2}$	&	0.0488	$\pm$	0.0003	&	0.1115	$\pm$	0.0009	\\
$f_{3}$	&	0.0089	$\pm$	0.0003	&	-0.4289	$\pm$	0.0048	\\
$f_{1}+f_{2}$&	0.0169	$\pm$	0.0003	&	0.1091	$\pm$	0.0025	\\
$2f_{1}$	&	0.0089	$\pm$	0.0003	&	0.3375	$\pm$	0.0047	\\
$f_{2}-f_{1}$	&	0.0040	$\pm$	0.0003	&	0.1372	$\pm$	0.0105	\\
$2f_{2}$	&	0.0082	$\pm$	0.0003	&	-0.3968	$\pm$	0.0052	\\

\\
\hline
\hline
\end{tabular}
\end{table}

\section{Mode Identification}

\begin{figure*}
\centering
\includegraphics[scale=0.88]{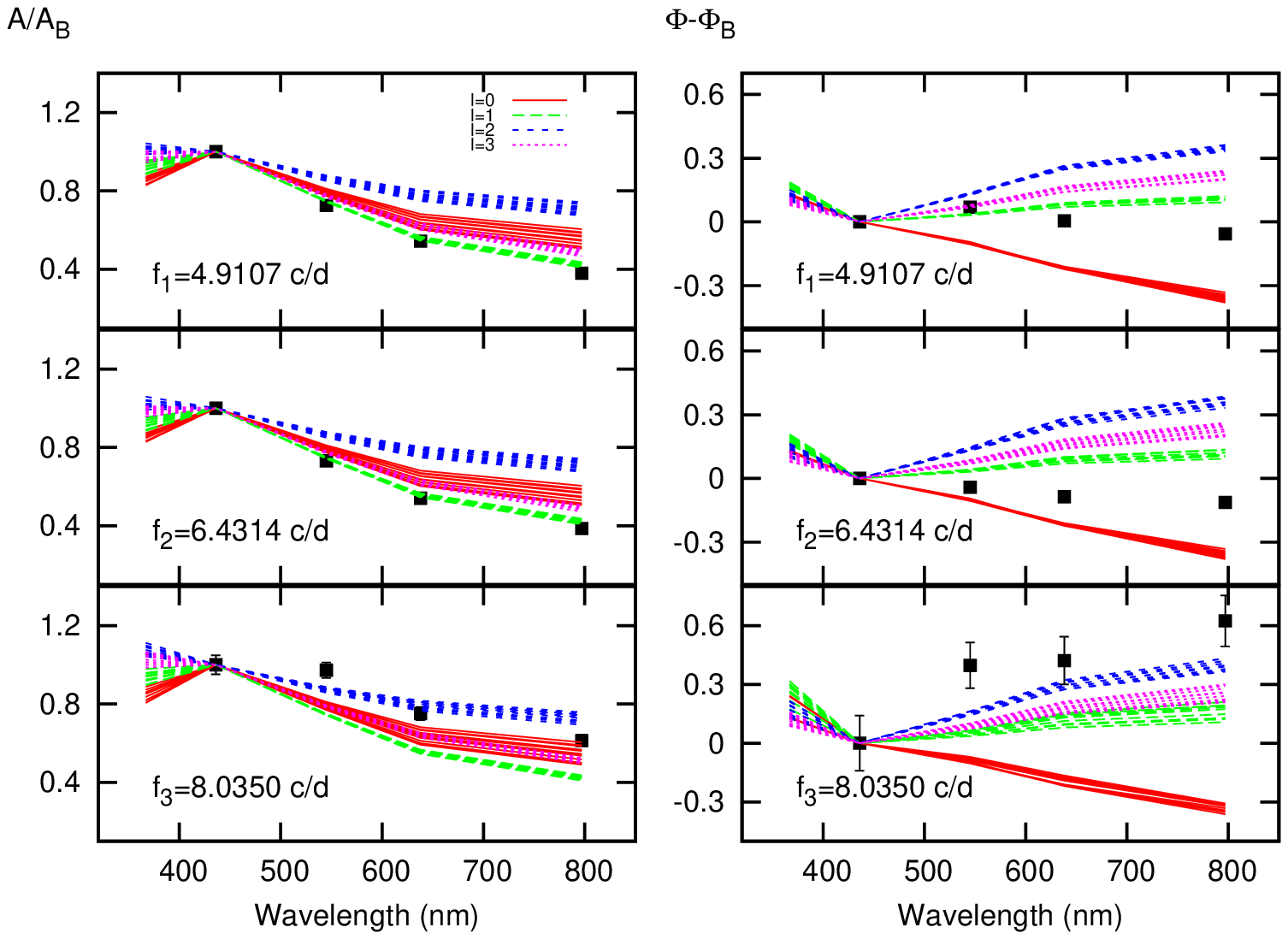}
\caption{Amplitude ratios (left panels) and phase differences (radians, right panels)
for $f_1, f_2$ and $f_3$.  The curves are from models with a range of stellar
parameters of the best estimated values with a range of stellar parameters of 
$6800 < T_{\rm eff} < 7200$\,K and $3.5 < \log g < 3.9$, and for different spherical harmonic
degrees, $l$ ($0 \leq l \leq3$).The red (solid) lines indicate $l = 0$, 
the green (long-dashed) $l =1$, the dark blue (dashed) $l=2$ the purple (dotted) $l =3$. 
The models computed for $M=1.90M_\odot$. The black squares represent the observed values
with their standard deviations for $f_1, f_2$ and $f_3$.}
\label{figmodeid}
\end{figure*}

Photometric mode identification is based on a comparison of observed amplitude 
ratios or phase differences with the computed values at different wavelengths. 
These differences are small and requires photometry of high accuracy.  The
best results are obtained with pulsations of high amplitude since the
relative errors are smaller. With this in mind, we only considered the
three independent modes with highest amplitudes in KIC\,6382916. 
The method followed here is described in a previous multi-site campaign reported 
by \cite{ulu13}. In order to calculate the amplitude ratios and phase differences 
in the Johnson/Cousins $BVRI$ system for a given spherical harmonic degree,
$l$, we made use of the {\tt FAMIAS} software package \citep{zim08}. This program 
uses pre-computed grids of non-adiabatic stellar models.  The calculations 
need to be restricted to models in the appropriate range of 
$T_{\rm eff}$, $\log g$, stellar mass and metallicity.

For KIC\,6382916,  we computed the amplitude ratios and phase differences in
the $BVRI$ bands for spherical harmonic degrees $0\le l\le 3$.  We adopted the 
stellar parameters $T_{\rm eff} =6950\pm100$\,K,  $\log g = 3.7\pm0.1$  and 
microturbulent velocity $\xi = 2$\,km\,s$^{-1}$. Using these parameters, the 
mass of KIC\,6382916 is estimated to be $M=1.93\pm0.27M_\odot$ from the
empirical relations in  \cite{tor10}. The amplitude ratios and phase 
differences depend on the stellar parameters.  Since the estimated
stellar parameters are not very precise, we decided to calculate these values 
over the range $6800 < T_{\rm eff} < 7200$\,K and $3.5 <\log g < 3.9$.

\begin{figure*}
\includegraphics[scale=0.8]{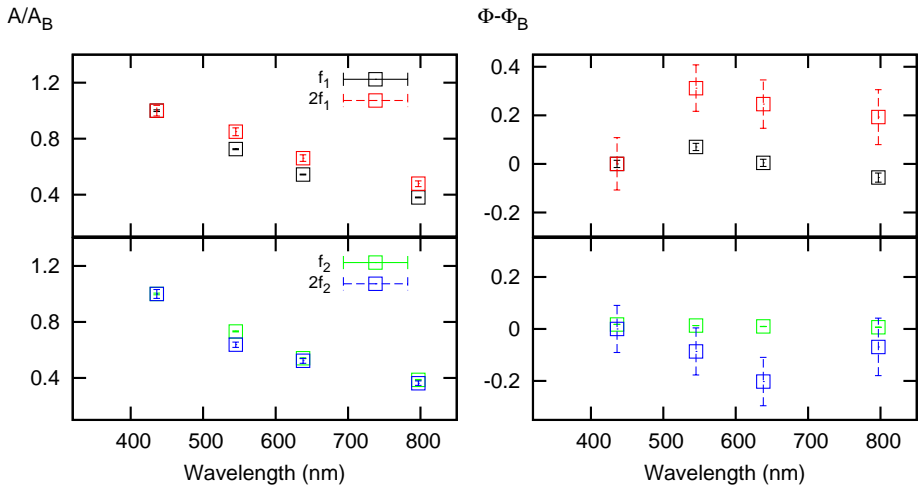}
\caption{Amplitude ratios (left panel) and phase differences (radians, right
panel) for $f_1$, $f_2$ and its harmonics, $2f_1, 2f_2$.}
\label{figharm}
\end{figure*}

In {\tt FAMIAS}, theoretical calculations are performed adopting a grid for 
$\delta$\,Scuti stars computed with the {\tt ATON} \citep{ven08} and {\tt MAD}  
\citep{mon07} codes. The uncertainity in the theoretical prediction of the amplitude ratios comes from the uncertainties of the parameters, amplitude, phase, effective temperature and   $\log g$ \citep{zim08} . The observed amplitude ratios and phase differences for $f_1$, $f_2$ and $f_3$ normalized to the $B$ filter are listed in 
Table\,\ref{tabampphase}.   Comparison with the calculated values are shown in 
Fig.\,\ref{figmodeid}.  Also, the relationship between the amplitude ratios, phases and these frequencies are shown in Fig. \ref{figharm}. 

As can be seen from Fig. \ref{figmodeid}, the observed amplitude ratios for  
$f_1$ and $f_2$ agree quite well if $l = 1$ for both modes.  We could not 
obtain a reliable mode identification for $f_3$ because of its low amplitude, 
but it appears to be either $l=0$ or $l=2$.  The phase differences $f_1$ agree 
quite well for $l=1$, though for $f_2$ they are intermediate between $l = 1$ 
and $l = 0$.  Overall, it seems as if $f_1$ and $f_2$ are both dipole modes.
The radial mode seems to be ruled out for both modes.

\begin{table}
\centering
\scriptsize
\caption{Amplitude ratios, $A/A_B$, and phase differences, $\phi - \phi_B$,
(radians) for the seven frequencies of highest
amplitude.}
\label{tabampphase}
\begin{tabular}{lrrr}
\hline
\hline
\multicolumn{1}{c}{Filters} &
\multicolumn{1}{c}{$A / A_B$} &
\multicolumn{1}{c}{ } &
\multicolumn{1}{c}{$\phi - \phi_B$} \\
\hline
 &   & $f_1$  &   \\
$B$	&	1.0000$\pm$0.0055	&		&   0.0000$\pm$0.0151  \\
$V$	&	0.7252$\pm$0.0037	&		&   0.0697$\pm$0.0151  \\
$R$	&	0.5433$\pm$0.0031	&		&   0.0044$\pm$0.0151  \\
$I$	&	0.3802$\pm$0.0028	&		&   -0.0565$\pm$0.0188  \\
&   & $2f_1$  &   \\					    		
$B$	&	1.0000$\pm$0.0380	&		&   0.0000$\pm$0.1081  \\
$V$	&	0.8495$\pm$0.0280	&		&   0.3123$\pm$0.0955  \\
$R$	&	0.6613$\pm$0.0240	&		&   0.2463$\pm$0.0993  \\
$I$	&	0.4785$\pm$0.0206	&		&   0.1929$\pm$0.1131  \\
&   & $f_2$  &   \\					    		
$B$	&	1.0000$\pm$0.0056	&		&   0.0000$\pm$0.0151  \\
$V$	&	0.7324$\pm$0.0037	&		&   -0.0415$\pm$0.0151  \\
$R$	&	0.5408$\pm$0.0032	&		&   -0.0861$\pm$0.0151  \\
$I$	&	0.3864$\pm$0.0028	&		&   -0.1125$\pm$0.0188  \\
&   & $2f_2$  &   \\					    		
$B$	&	1.0000$\pm$0.0313	&		&   0.0000$\pm$0.0905  \\
$V$	&	0.6372$\pm$0.0194	&		&   -0.0867$\pm$0.0905  \\
$R$	&	0.5221$\pm$0.0176	&		&   -0.2023$\pm$0.0930  \\
$I$	&	0.3628$\pm$0.0155	&		&   -0.0691$\pm$0.1106  \\
&   & $f_3$  &   \\					    		
$B$	&	1.0000$\pm$0.0488	&		&   0.0000$\pm$0.1407  \\
$V$	&	0.9724$\pm$0.0394	&		&   0.3971$\pm$0.1169  \\
$R$	&	0.7517$\pm$0.0332	&		&   0.4216$\pm$0.1219  \\
$I$	&	0.6138$\pm$0.0296	&		&   0.6239$\pm$0.1307  \\
\hline
\hline
\end{tabular}
\end{table}

\section{Conclusions}

Analysis of short-cadence {\it Kepler} photometry was used to determine the
pulsation frequencies in KIC\,6382916.  We find two large-amplitude independent
modes with frequencies $f_1$ = 4.9107 and $f_2$ =6.43137\,d$^{-1}$ as
previously reported by Pigulski et al. (2009) from ground-based observations.
We found a third independent frequency at $f_3 = 8.03506$\,d$^{-1}$ which
has a much lower amplitude.  The frequency spectrum is dominated by $f_1$ and 
$f_2$ and their harmonics and combination frequencies.

In HADS stars the modes of highest amplitude are generally radial modes
since in many stars the period ratio $P_1/P_0$ of first overtone to
fundamental radial modes is close to the expected value $0.77< P_1/P_0<
0.78$.  The stars lie on a curve defined by $P_1/P_0$ as a function of $\log
P_0$ which is called the Petersen diagram \citep{pet73}.  The period ratio
depends on metallicity, rotation and chemical abundance \citep{l08, su07}. 
In KIC\,6382916 the period ratio for the two modes of highest amplitude is
$f_1$/$f_2$=0.763 which differs significantly from the expected period ratio
for fundamental and first overtone radial modes.  It therefore seems that at
least one of the two modes is probably a nonradial mode.  This may not be
too surprising since recent studies suggest that the radial mode need not be
present in all HADS stars \citep{por11}. 

To determine the spherical harmonic degree of the three independent modes in
KIC\,6382916 we first of all need to know the stellar parameters as
accurately as possible.  For this purpose we obtained high-dispersion spectra 
and estimated an effective temperature $T_{\rm eff} =6950\pm100$\,K and 
$\log g = 3.7\pm0.1$ by matching the Balmer line profiles with profiles
calculated from synthetic spectra.  From the effective temperature and
surface gravity we estimate the mass as $M=1.93\pm0.27M_\odot$.

Mode identification is best done by comparing the relative amplitude ratios 
and/or phase differences in different photometric wavebands for the required 
mode with the calculated values.  For this purpose we initiated a multi-site 
photometric campaign to observe KIC\,6382916 in the $BVRI$ bands.  We used the 
frequencies derived from the {\it Kepler} data as fixed values and fitted a 
truncated Fourier series to the data from each wave band using the seven 
frequencies of highest amplitude.  The resulting amplitude and phases were 
used to construct the relative amplitudes and phase differences, normalized to 
the $B$ band, for $f_1$, $f_2$ and $f_3$.  We then used the {\tt FAMIAS} software 
package \citep{zim08} to calculated amplitude ratios and phase differences for 
models with stellar parameters approximately corresponding to the spectroscopic 
values mentioned above.  It turns out that $f_1$ and $f_2$ are both dipole
($l = 1$) modes and that $f_3$ is either $l = 0$ or $l = 2$.

If the identification for both $f_1$ and $f_2$ is correct, it calls to mind
the case of 1\,Mon \citep{bal80,bal01}. In this star there are three modes:
a central radial mode flanked by two $l = 1$ modes.  In any case it seems
that we need to be cautious in attributing the high-amplitude modes in HADS
stars as radial modes, though this is probably true in the majority of
cases.

It is important to compare the observed frequencies of $f_1$ and $f_2$ with
model frequencies for $l = 1$.  However, we are faced with the problem that
rotation strongly modifies the frequencies of the $l=1$ modes.  We therefore
need to estimate the possible frequency shift owing to this effect which
requires knowledge of the rotation frequency.   In some $\delta$\,Sct stars, 
it is possible to detect the rotational frequency directly from the periodogram 
of the {\it Kepler} data by looking for the presence of a low-frequency peak 
and its harmonic.  The presence of an harmonic is an indicator that the peak 
is due to a starspot and hence the frequency is the rotational frequency 
\citep{bal13}. Unfortunately, no such peak is visible in KIC\,6382916.

The alternative is to estimate the rotational frequency from the projected
rotational velocity  $v\sin\,i = 50 \pm 10$\,km\,s$^{-1}$ and the stellar
radius.  The stellar radius is estimated to be about 3.69$\pm$0.14\,$R_\odot$ 
using the relationships by \cite{tor10}.  If we assume that the star is 
roughly equator-on, the equatorial rotational velocity will probably be about
50 km\,s$^{-1}$ and the rotation frequency around 0.27\,d$^{-1}$.  If the 
star is equator-on, only sectorial modes will be visible and the frequency
shift will roughly be the same as the rotation frequency.  Thus we may
expect the frequencies in the non-rotating frame to be in the range 
$4.6 < f_1 < 5.2$ and $6.1 < f_2 < 6.7$. 

We can compare these frequencies with frequencies calculated from non-rotating
models. For this purpose, models of $\delta$~Scuti stars with masses in the range
$1.3 < M/M_\odot < 2.5$ were constructed using the Warsaw - New Jersey code
\citep{pac70}. These models use OPAL opacities, no core overshoot and a
mixing length, $\alpha = 1.0$.  Pulsation frequencies for each model were obtained
using the {\tt NADROT} code \citep{dzi77}. It turns out that all modes with frequencies 
less than about 6.1\,d$^{-1}$ are stable. One may argue that the rotation frequency is
much larger and that the rotation correction would then bring $f_1$ closer to the
this value.  This argument cannot be correct since this would imply a nearly
pole-on orientation, in which case sectorial modes are no longer visible. 
We are thus faced with the problem that current models are unable to account
for driving at the low frequency of $f_1$. Models where $f_1$  
and $f_2$ are both present (but with $f_1$ stable) all have $6800 < 
T_{\rm eff} < 7200$\,K and $3.78 <\log g < 3.88$ which is roughly in the 
range of the values determined from spectroscopy. In the model, frequencies 
in the range of $f_1$ and $f_2$ are mixed p and g modes of
high radial order, but this does not mean that the observed modes are also
of this kind since no model is capable of driving these frequencies. The
lowest frequency of a p mode for models in the above parameter range is about
8.6 \,d$^{-1}$. They all appear to be g modes of high order.  
The lowest frequency of a p mode for models in the
above parameter range is about 8.6\,d$^{-1}$.

In conclusion, we find that the two modes of highest amplitude in
KIC\,6382916 are not only dipole modes, contrary to expectation, but are
high-order g modes which are predicted to be stable in the models.  Unless
the derived stellar parameters are grossly in error, this would imply a
problem in our understanding of pulsational driving in $\delta$~Scuti stars. 
We know that there is a problem with the models in that the observed  range 
of instability is wider than the calculated range for $\delta$~Sct stars 
\citep{bal11}.  It appears that the presence of two high-amplitude g modes
in KIC\,6382916 is further confirmation of this problem.

\section*{Acknowledgments}

The authors acknowledge the whole {\it Kepler} team for providing the 
unprecedented data sets that makes these results possible.
This paper includes data collected by the Kepler mission. 
Funding for the {\it Kepler} mission is provided by the NASA Science Mission directorate.  
CU sincerely thanks the South African National Research Foundation (NRF) for the award of 
innovation post doctoral fellowship, Grant No. 73446. BU is supported by the 
project numbered HDYF-051. TG would like to thank NRF Equipment-Related 
Mobility Grant-2011 for travelling to carry out the photometric 
observations.  LAB thanks the South African National Research Foundation 
and the South African Astronomical Observatory for generous financial 
support.  IS and II gratefully acknowledge the partial support from 
Bulgarian NSF under grant DO 02-85. DD acknowledges for the support of 
grants DO 02-362 and DDVU 02/40-2010 of Bulgarian NSF.  HAK acknowledges 
Carlos Vargas-Alvarez, Michael J. Lundquist, Garrett Long, Jessie C. Runnoe, 
Earl S. Wood, Michael J. Alexander for helping with the observations at WIRO.
LFM acknowledges financial support from the UNAM under grant PAPIIT  104612
and from CONACYT by the way of grant 118611.  MD, AC and DC are supported 
by grants provided by the European Union, the Autonomous Region of the Aosta 
Valley and the Italian Department for Work, Health and Pensions. The OAVdA 
is supported by the Regional Government of the Aosta Valley, the Town 
Municipality of Nus and the Mont Emilius Community. TEP aknowledges support 
from the National Research Foundation of South Africa. This study made use 
of {\tt IRAF} Data Reduction and Analysis System and the Vienna Atomic Line 
Data Base (VALD) services.  The authors thank Dr Zima for providing the 
{\tt FAMIAS} code.

\label{lastpage}

\end{document}